\documentclass[a4paper,14pt]{extarticle}
\usepackage{amsmath,amssymb,amsfonts,amsthm,upgreek} 
\usepackage[left=1.5cm,right=1.5cm,top=2cm,bottom=2cm,bindingoffset=0cm]{geometry} 
\usepackage{indentfirst} 
\usepackage{setspace}
\usepackage{hyperref}
\hypersetup{colorlinks=true, linktoc=all, linkcolor=black, urlcolor=black, citecolor=black}

\usepackage{graphicx}
\usepackage{subcaption}
\usepackage{svg}
\graphicspath{{pics/}}
\DeclareGraphicsExtensions{.pdf,.png,.jpg,.eps}
\usepackage{overpic}

\usepackage[
backend=biber,
style=numeric,
sorting=none
]{biblatex}
\title{Rotation-induced phase transition in planar continuous helicity gas}
\author{M.E. Malev, D.S. Kaparulin and N.N. Levin}

\begin{document}

\maketitle
\begin{abstract}
    In this paper, we investigate the thermodynamics of an ideal gas of classical particles with continuous helicity in three-dimensional Minkowski space. Using the one-particle distribution function for a particle with continuous helicity, we obtain expressions for the chemical potential, angular momentum, and entropy of the gas. We show that such a system placed in a rotating container can be in two phases: rotating and non-rotating. We describe the conditions for the phase transition and examine the phase diagram of the gas. It is found that at low angular velocities, the rotating phase can exist only in the form of metastable states. A feature of metastable states with low angular velocities is a negative angular momentum.
\end{abstract}

\section{Introduction} \label{sec:I}

Modern theoretical physics describes elementary particles in terms of unitary irreducible representations of the Poincaré group. The classification of these representations was developed by Wiegner and includes both massive and massless particles, as well as particles of continuous helicity (continuous spin) \cite{Wiegner1948}. Continuous spin particles occupy an intermediate position between massive and massless particles. Despite having zero rest mass, they possess the same number of degrees of freedom as massive particles \cite{Lyakhovich:1996}. The practical significance of such a model is that, at high temperatures, it can imitate the behaviour of massive particles. The concept of a particle with continuous helicity can be generalised to the case of higher or lower dimensions \cite{Lyakhovich:1996}, \cite{Buchbinder:2020}, \cite{LYAKHOVICH_1996}. The model of a particle with continuous helicity in three-dimensional Minkowski space was developed in the paper \cite{Gorbunov_1997}. It is known that the trajectories of a particle with continuous helicity are confined to a parabolic cylinder within Minkowski space \cite{Kaparulin2022}. The statistical mechanics of a rotating ideal gas of particles with continuous helicity was studied in the paper \cite{Kaparulin2023}. It was demonstrated that the model displays the effect of partial non-rotation, whereby the regular rotation of the gas container is possible only in one direction depending on the sign of the helicity. In addition, chiral effects are observed in the model, expressed as anisotropy of the momentum distribution in the presence of external rotation. It was also found that the transition to a regular rotation state occurs abruptly and is accompanied by a discontinuous change in angular momentum. The phenomenon has no direct analogues in the gas model of massive or massless spin particles.

In this study, a detailed examination is conducted on the complex phenomena that occur during the rotation of a two-dimensional ideal gas composed of spin particles with continuous helicity, confined within a circular vessel. Within the context of thermodynamic analysis in the model of a rotating gas of spin particles exhibiting chiral effects, references such as \cite{Gibbs1902}, \cite{FUKUSHIMA2019167}, \cite{Fukushima2013}, \cite{Vilenkin1980} are drawn upon. It is demonstrated that this process can be described by a two-phase model, where the entire gas may exist in one of two phases: rotating and non-rotating. The rotating phase possesses a nonzero angular momentum and participates in the rotation alongside the vessel. The non-rotating phase has zero angular momentum and does not partake in the rotation. When the angular velocity is below a certain critical threshold, the non-rotating phase is the most stable; above this threshold, the rotating phase becomes more stable. The phase transition is accompanied by a discontinuous change in the entropy and angular momentum of the gas, classifying it as a first-order phase transition. The critical angular velocity at which the phase transition occurs to be depended on the gas temperature, the size of the vessel, and the helicity of the particle. Higher values of helicity increase the critical angular velocity. At angular velocities lower than the critical value, the rotating phase exists in the form of a metastable state. A notable feature of this state is the negative value of the moment of inertia at the low angular velocity values. A similar effect has recently been observed in the gas of ultra-relativistic massive particles as reported in \cite{BRAGUTA2024}.

% М: Спорный абзац, не уверен, нужен ли он
The structure of the work is as follows. The first section describes the model of a particle with continuous helicity under consideration and presents the results obtained for it. An expression for the partition function of a single particle is derived, and convenient dimensionless parameters are introduced and explained. This section also provides a description of the chemical potential of a gas consisting of $N$ particles, along with the derivation of expressions for the entropy and angular momentum of the gas. In the second section, a more detailed analysis of the chemical potential is conducted in the context of studying the phase transition between the rotating and non-rotating phases. The stability conditions for the rotating and non-rotating phases are described, as well as the conditions for the existence of the phase transition. The magnitudes of the entropy and angular momentum gaps are calculated, and a phase diagram is constructed and analyzed. The conclusion summarizes the results obtained.

\section{Thermodynamic potential}\label{sec:CP}

We consider a classical ideal gas of continuous helicity particles in rotating circular container in $3d$ Minkowski space-time. 

The state of a single particle is determined by the vector of linear $p_\mu$ and total angular momentum $j_\mu$, $\mu=0,1,2,$ subjected to the mass shell conditions
\begin{equation}\label{ms}
    (p,p)=0\,,\qquad (p,j)=\sigma\,.
\end{equation}
The brackets denote the scalar product with respect to the Minkowski metric. We use the mostly positive convention for the signature of the metric. The quantity $\sigma$ denotes the helicity of the particle. Without loss of generality, we assume $\sigma>0$. The case of negative helicity can be considered after spatial reflection $p\to-p$. The conditions (\ref{ms}) can be considered as the extreme case of a zero-mass particle with infinite spin. In this case we have the estimate $\sigma=mcs\hbar$, where $s$ is the spin and a $m$ is the particle mass. The quantities $c$, $\hbar$ are the speed of light and the Plank constant. The relationship between helicity, mass and spin allows the continuous helicity model to be used to study hot gases of real particles with mass and spin. 

In the article \cite{Kaparulin2023} it was shown that the one-particle phase space can be parameterised by its spatial position $\boldsymbol{x}=(x^i,i=1,2),$ and the spatial part of the linear momentum $\boldsymbol{p}=(p^i,i=1,2)$. The space-time momentum and the total angular momentum are expressed as follows:
\begin{equation}
    p^\mu=(|\boldsymbol{p}|c, p^i)\,,\qquad j^\mu=\left(x^1p^2-x^2p^1+\frac{1}{|\boldsymbol{p}|}\sigma, p^i c\left(x^1-x^2\right)+\frac{1}{|\boldsymbol{p}|^2}\sigma p^i\right)\,.
\end{equation}
Here, $|\boldsymbol{p}|$ denotes the norm of the spatial momentum vector $\boldsymbol{p}$. In the same work, the one-particle classical distribution function $f_0$ for continuous helicity particles was derived using the Boltzmann statistics,
\begin{equation}\label{f0}
    f_0(x^i,\Tilde{p}^i) = \frac{1}{Z_0} \exp\left(-\frac{|\boldsymbol{\Tilde{p}}| \sqrt{c^2-\omega^2r^2}}{\theta}-\frac{\omega\sigma(|\boldsymbol{\Tilde{p}}| c-\omega(x^1\Tilde{p}^2-x^2\Tilde{p}^1))}{2\theta |\boldsymbol{\Tilde{p}}|^2\sqrt{c^2-\omega^2r^2}}\right)\,.
\end{equation}
Here $\Tilde{p}^i$ are the components of the particle momentum, measured in the rotating frame (i.e. by the "rotating" observer) \cite{Kaparulin2023}. In what follows we use only the $\Tilde{p}$ vector of momentum so for the sake of convenience we note it simply as $p$, without the tilde. The quantity $r=\sqrt{(x^1)^2+(x^2)^2}$ denotes the distance from the origin. To avoid superluminal rotation, we assume that the radius of the circle is sufficiently small, $R<\omega/c$. The normalisation multiplier determines the one-particle partition function $Z_0$\,,
\begin{equation}\label{Z0}
    Z_0 = \int\exp\left(-\frac{p \sqrt{c^2-\omega^2r^2}}{\theta}-\frac{\omega\sigma(pc-\omega(x_1p_2-x_2p_1))}{2\theta p^2\sqrt{c^2-\omega^2r^2}}\right)dx^1dx^2dp^1dp^2\,.
\end{equation}
If $\omega$ and $\sigma$ have different sings, the integral diverges. It can be interpreted as the phenomenon of partial non-rotation of the classical gas of continuous helicity particles. It means that only a single direction of rotation is available for the macroscopic system. The direction of the rotation depends on the sign of helicity of the particles. If $\sigma>0$ the system can rotate only counterclockwise ($\omega\geq0$). If $\sigma<0$, then the only possible direction is clockwise ($\omega\leq0$). Both cases are equivalent. We assume that $\sigma>0$.

The analytic expression for integral in (\ref{Z0}) is long and not very informative. It is convenient to consider slow rotating gas. Expanding this function to the second order in $\omega$ gives the following estimate:
\begin{equation}\label{Z0-xA}
    Z_0 = \frac{\theta^2V}{2\pi c^2 \hbar^2}\left[1-x-\frac{x^2}{2}(\ln Ax+1)+o(x^2)\right]\,.
\end{equation} 
Here $V=\pi R^2$ is the volume of the container, $R$ is its radius and $\gamma=0.577\ldots$ is Euler's constant. The expression has a clear physical meaning: the common factor determines the (one-particle) partition function of the non-rotating gas, and the expression in brackets accounts for the rotational corrections up to second order in angular velocity. The rotational correction depends on two dimensionless parameters, which we denote as $x$ and $A$:
\begin{equation}\label{xk}
    x = \frac{c\omega\sigma}{2\theta^2}\,, \qquad \ln A = 2\gamma-\frac{1}{2}-\left(\frac{2\theta^2R}{c^2\sigma}\right)^2\,.
\end{equation}
The first parameter $x$ can be considered as a dimensionless angular velocity. The second parameter $A$ takes into account the geometry of the system. In general, $x$ and $A$ can vary on a very large scale, but in the high temperature limit they should be small quantities. Considering a gas of continuous helicity particles as an ultra-relativistic limit of ordinary gas of massive particles with electron mass and spin $s$, we can estimate $\sigma=m_ec\hbar s \approx2.88\cdot 10^{-56} s \text{ J$\cdot$kg$\cdot$m}$\,, and
\begin{equation}\label{xk-num}
    x \approx 0.0226\cdot\frac{\omega s}{T^2}\,,\qquad \ln A \approx 0.154-3.34\cdot 10^{-10}\frac{R^2T^4}{s^2}
\end{equation}

Here, the angular velocity is measured in Hertz, temperature in Kelvins, and $s$ is a dimensionless quantity. In the range of temperatures exceeding the rest energy of the electron $T=5.92\cdot 10^{9}$ K, quantity $x$ is small for reasonable values of angular velocities. As for the quantity $A$, it is small for not too small systems.
For example, in case (\ref{xk-num}) and for $s=1$, $\omega=10^5$, $T=10^{11}$ K, $R=1$ m, we obtain $x\approx0.0026\cdot 10^{-17}$ and $\ln A \approx -3.34\cdot 10^{34}$.

Now, we can consider the gas consisting of $N$ particles. The grand thermodynamic potential of the rotating gas of classical continuous helicity particles is (see \cite{LandauV})
\begin{equation}\label{Omg}
    \Omega = -\theta \ln \left(\sum_{N=0}^\infty\frac{1}{N!}e^\frac{\mu N}{\theta}Z_0^N\right) = -\theta \ln \left(\exp\left(Z_0 e^\frac{\mu}{\theta}\right)\right) = -\theta Z_0 e^\frac{\mu}{\theta} \,,
\end{equation}
with $\mu$ being the chemical potential. The thermodynamic potential is a function of variables $\theta$, $\omega$, $\mu$. The differential of the thermodynamic potential reads
\begin{equation}\label{dOmega}
    d\Omega=-Sd\theta-Jd\omega -Nd\mu.
\end{equation}
Here, $S$ is entropy, $J$ is the total angular momentum. The equation (\ref{dOmega}) gives the expressions for $S$, $J$ and $N$:
\begin{equation}\label{SJN}
    S=-\frac{\partial\Omega}{\partial \theta}\,,\qquad 
    J=-\frac{\partial\Omega}{\partial \omega}\,, \qquad 
    N=-\frac{\partial\Omega}{\partial \mu}\,.
\end{equation}
In what follows, we use only the specific values for the entropy and angular momentum: $S/N$, $J/N$. For simplicity, we note them the same way, $S$ for the specific entropy and $J$ for the specific angular momentum.

The last equality in (\ref{SJN}) allows us to express chemical potential as the function of number of particles, temperature and volume:
\begin{equation}\label{mu}
    \mu = -\theta\ln \frac{\theta^2V}{2\pi c^2 \hbar^2N}+\theta\left[x+\frac{1}{2}x^2\ln (Ax)\right]\,,
\end{equation}
where the notation (\ref{xk}) is used. The first term defines the chemical potential of the non-rotating gas, and the second term defines the rotational correction. Differentiating $\Omega$ with respect to the temperature and using the equation of state (\ref{mu}), we find the specific entropy of the gas,
\begin{equation}\label{S}
    S=\ln \frac{\theta^2V}{2\pi c^2 \hbar^2N}+ 3+ x + x^2\left(\frac{3}{2}\ln Ax + 2\ln A + 4\gamma\right)\,.
\end{equation}
By construction, entropy includes the contribution from non-rotating term, and rotational correction, which is proportional to the angular velocity $x$. The specific total angular momentum of the gas reads 
\begin{equation}\label{J}
    J = -\frac{c\sigma}{2\theta}\left(1+\frac{1}{2}x+x\ln Ax\right)\,.
\end{equation}

The thermodynamic parameters (\ref{S}) and (\ref{J}) demonstrate an interesting feature. While a small increase in angular velocity causes a small increase in entropy, the angular momentum undergoes a discontinuous change at the transition from the rotating to the non-rotating state:
\begin{equation}\label{J0}
    J\Big|_{\omega=0}=0\,,\qquad \lim_{\omega\to +0} J=-\frac{c\sigma}{2\theta}\,.
\end{equation}
% переписать этот абзац по человечески
The first equality expresses the fact that for a non-rotating gas both the momenta and the spins have an isotropic distribution. The second equality follows from the formula (\ref{J}). To explain this discontinuity of angular momentum, we should assume that at zero angular velocity the gas can exist in two phases. The first is a non-rotating one, which has zero angular momentum, and the second is a rotating one, which can rotate in one direction. The formula (\ref{Omg}) describes both of these cases. The discontinuous change in angular momentum is explained by a first order phase transition between the two phases. The rotational instability in the model was recently noted in \cite{BRAGUTA2024} in the context of the study of rotating gluon plasma. The existence of a phase transition between the rotating and non-rotating states was first noted in liquid helium, see for example \cite{Hall-1960}. We will discuss this phenomenon in more detail below. 
% М: возможно ссылку на глюонную плазму стоит поставить ближе к описанию отрицательного углового момента, но вроде и тут норм.

\section{Phase transition and metastable states}\label{sec:PT}

The study of the chemical potential provides the most relevant framework for describing phase transition phenomena. If the chemical potential of two phases is the same, the phases are in equilibrium \cite{kubo}. Otherwise, the phase with the lower chemical potential is more stable. In the case of our model, the stability of the phases is determined by the rotational correction to the chemical potential
\begin{equation}\label{mu_r_A}
    \mu_r(x,\theta)\equiv\mu(x,\theta)-\mu(0,\theta)= \theta \left(x+\frac{x^2}{2}\ln Ax\right)\,.
\end{equation}
If the correction (\ref{mu_r_A}) is positive, the non-rotating phase is more stable. If it is negative, the rotating phase is more stable. The condition for phase equilibrium is therefore
\begin{equation}\label{pe}
    \mu_r(x,\theta)=0\,.   
\end{equation}
For a given temperature it can be considered as the equation that determines the angular velocity of the phase transition. Depending on the value of the parameter $A$ the equation (\ref{pe}) can have one, two or three roots. 

If $\ln A<-1$, it has three roots, which we denote as $x_0,\,x_1$ and $x_2$. The lowest root $x_0$ is zero, and the other two are expressed in terms of the Lambert function branches $W_0$ and $W_{-1}$,
\begin{equation}\label{x}
    \qquad x_1=-\frac{2}{W_{-1}(-2A)}\,,\qquad x_2=-\frac{2}{W_{0}(-2A)}\,.
\end{equation}
The first point of equilibrium $x_0$ corresponds to a non-rotating system. This is not surprising, as both phases have no rotation. The solution $x_1$, which lies between $x_0$ and $x_2$, determines the value of the angular velocity at the phase transition. The rotational correction (\ref{mu_r_A}) is positive for $0<x<x_1$ and negative for $x_1<x<x_2$. Thus, the rotational phase is less stable in the low angular velocity range. It becomes more stable for $x>x_1$. The third root $x_2$ is many orders greater than $x_1$. The principal branch $W_0(z)$ of the Lambert function never corresponds to small angular velocities, see \cite{onLambert}.
%The principal branch $W_0(z)$ of the Lambert function can be expanded near the point $z=0$ in the power series with the radius of convergence $1/e$, see \cite{onLambert}. Thus, the lesser is the parameter $A$, the larger the value of $x_2$ becomes. 
It is not relevant for small $x$, being the subject of the slow-rotating gas model. Thus, we consider the third root to be an artifact of the slow rotation approximation.

\begin{figure}[h]
    \centering
    \includegraphics[scale=0.6]{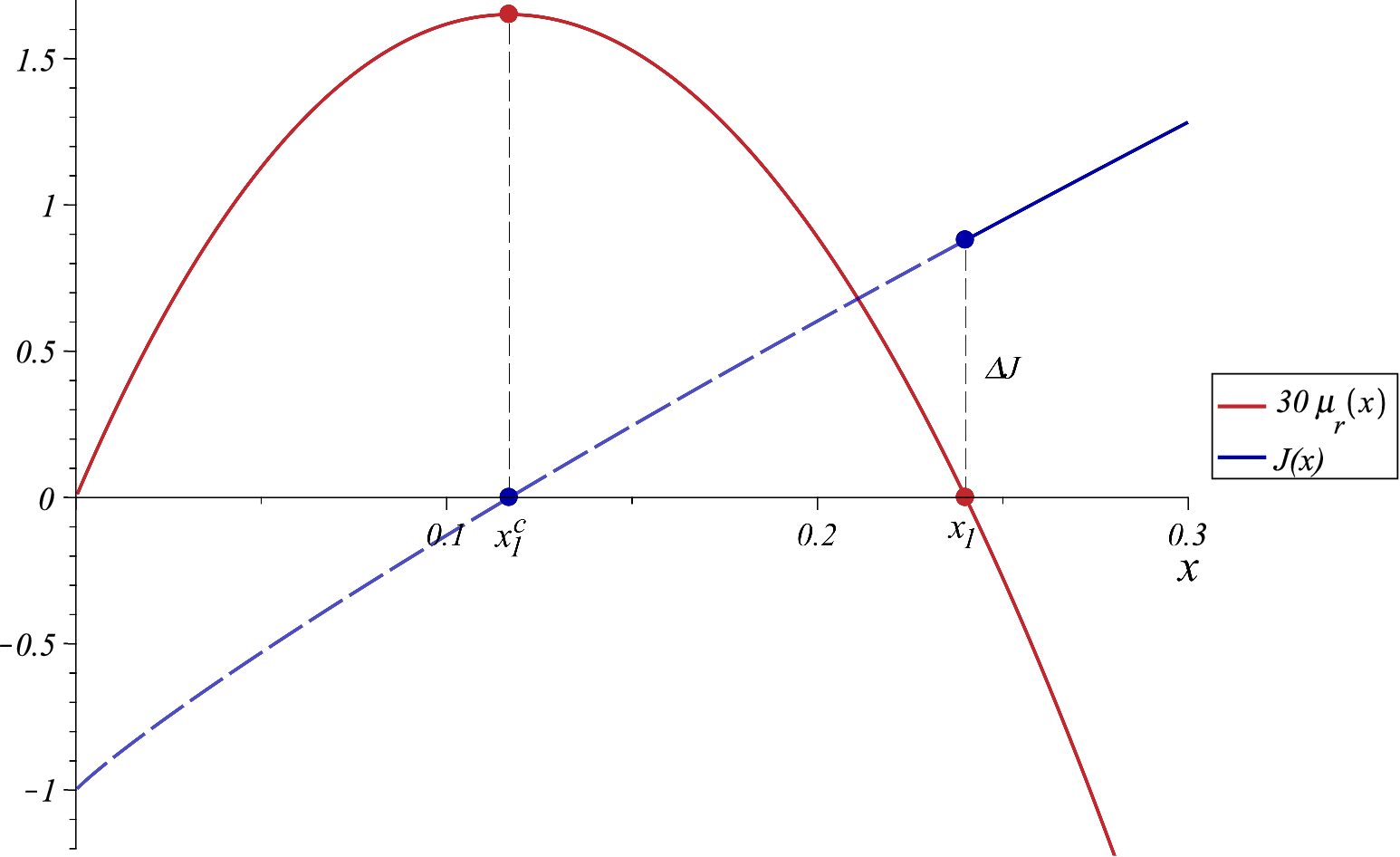}
    \caption{Plot of the $\mu_r (x)$ (red line) and $J(x)$ (blue line) functions with $\theta = 1$, $A = 10^{-3}$. The function $\mu_r$ is multiplied by 30 to make the plot easier to read.}
    \label{fig:mu+j}
\end{figure}

The general behaviour of the correction $\mu_r$ in the $\ln A <-1$ case is shown on the figure \ref{fig:mu+j} (red line). It is positive between the points $x_0$ and $x_1$ and negative after the $x_1$ point all the way up to the $x_2$ point. For the given parameter $A=10^{-3}$ the coordinate of the third zero is $x_2\approx997.998$. It is located far beyond the range of reasonable angular velocities and isn't shown on the figure.

If $\ln A \geq -1$, the chemical potential $\mu$ becomes a strictly increasing function of the angular velocity. The equilibrium between the phases is only possible in the absence of rotation, $x=0$. In this case, the system cannot rotate in principle. This corresponds to a some sort of non-rotational regime. The requirement of rotation implies restrictions on the model parameters. It has the form:
\begin{equation}
    -1 > 2\gamma-\frac{1}{2}-\left(\frac{2\theta^2R}{c^2\sigma}\right)^2\,.
\end{equation}
If the continuous helicity gas is the ultra-hot gas of massive rotating particles, we get a relation between the size of the system and the spin of the particles:
\begin{equation}\label{restr}
    R > \frac{mc^3\hbar\sqrt{2\gamma+1/2}}{2\theta^2}\, s\,.
\end{equation}
This is not a strong constraint, because we assume that the particles are ultra-relativistic. However, it gives us two interesting results. First, if the spin is fixed, the formula (\ref{restr}) gives the lower limit for the radius of the system. In the case of $\theta = mc^2$ this limit becomes proportional to the reduced Compton wavelength:
\begin{equation}
    R > \frac{s\sqrt{2\gamma+1/2}}{2}\, \Bar{\lambda}_C\,.
\end{equation}
It means that the system must be larger than the scale at which quantum effects become significant. This confirms the self-consistency of our model, which is classical. 

Second, if the spin is not limited from above, the mass of the particle must be zero $m \to 0$. This observation has similarities with the higher-spin models, where the interacting theory for massless particles is well known. Massive higher spin models require a more cautious approach (for example, see \cite{PonomarevHigherSpins}). The formula (\ref{restr}) does not prevent the gas from rotating if the spectrum of masses and spins has a limited helicity $ms$ for all interacting particles. 
% М: можно сделать ссылку на что-нибудь про высшие спины, например, https://arxiv.org/abs/2206.15385 либо взять источники отсюда https://journals.aps.org/prl/abstract/10.1103/PhysRevLett.129.241601

The total angular momentum of the gas is given by the formula (\ref{J}). The critical points of $\mu_r$ determine the zero angular momentum states. If $\ln A<-1$ there are two critical points:
\begin{equation}\label{mu-extr}
    x^{c}_1 = -\frac{1}{W_{-1}\left(-A\sqrt{e}\right)}\,, \qquad x^{c}_2 = -\frac{1}{W_{0}\left(-A\sqrt{e}\right)}\,.
\end{equation}
The angular momentum is negative for $0<x<x^c_1,\, x^c_2<x<\infty$ and positive for $x^c_1<x<x^c_2$. A simple analysis shows that the first critical point is in the range $0<x^c_1<x_1$ and is a small value. The second solution is a big quantity, and it does not seem to be relevant in the context of the study of slowly rotating gas. Finally, the derivative of the angular momentum determines the specific rotational susceptibility of the gas,
\begin{equation}\label{chi}
    \chi \equiv \frac{\partial J}{\partial\omega} = 2\ln Ax +3 \,.
\end{equation}
The susceptibility function has a zero, $x^i=1/A e^{3/2}$, and is positive in the range $0<x<x^i_1$\,. This ensures the thermodynamic stability of the rotating phase, at least at low angular velocities. The analysis also shows that $x^c_1<x_1<x^i$, so the rotating phase is stable at the phase transition point.

The results of the analysis confirm that in the presence of external rotation the continuous helicity gas can be in two different phases: rotating and non-rotating. There are two cases of phase equilibrium. The first is the absence of rotation. The second is the phase equilibrium curve determined by the root $x_1$ (\ref{x}) of the rotational correction to the chemical potential. % М: может лучше сказать про "положение на кривой", а не про саму кривую?
For low angular velocities ($0<x<x_1$) the non-rotating phase is more stable. This also seems to be true for the point $x=0$. Indeed, angular momentum fluctuations increase entropy, making the transition to the non-rotating state energetically favourable. As for the rotating phase, the positive specific rotational susceptibility ensures that the rotating state does not disappear in the presence of small fluctuations. It is therefore observable. On the other hand, the correction $\mu_r$ is positive for $0<x<x_1$, so the rotating phase is less stable than the non-rotating one. We conclude that this is a metastable state. The peculiarity of this state is the negative value of the angular momentum for low angular velocities ($x<x^c_1$). It can be interpreted as the negative moment of inertia of the system. % М: вот сюда можно добавить ссылку про глюонную плазму, например (см комментарий выше)
The existence of such states does not contradict the usual understanding of rotation because they are metastable. The phase transition between the non-rotating and rotating phases occurs at the angular velocity $x=x_1$. This is a relatively large number, so extreme conditions are required to observe this phase transition. For $x>x_1$ the rotating phase is the most stable. As for the non-rotating phase, there is no reasonable way to preserve it for a gas in a rotating cylinder. Thus, this phase must be unobservable, and the rotating phase should be the only possible state for a rapidly rotating gas.

The phase transitions for $x=0$ and $x=x_1$ involve discontinuous changes in entropy and angular momentum. Consider the case of an increasing angular velocity $x$. As it approaches the point $x_1$ from the left ($x\to x_1-0$), the total angular momentum is zero and its entropy is given by the expression (\ref{S}) for $x=0$. At the same time, the angular momentum and entropy of the rotating phase are given by (\ref{S}), (\ref{J}) for $x=x_1$. The change of the thermodynamic quantities at the phase transition is given by
\begin{equation}\label{dS}
    \Delta S=S(x_1,\theta)-S(0,\theta)=\frac{4}{W_{-1}(-2A)}\left(1+\frac{2(\ln A +2\gamma)}{W_{-1}(-2A)}\right)\,;
\end{equation}
\begin{equation}\label{dJ}
    \Delta J=J(x_1,\theta)=\frac{c\sigma}{2\theta} \left(1+\frac{1}{W_{-1}(-2A)}\right)\,.
\end{equation}
This is an example of a typical first order transition as it involves a latent heat $\theta\Delta S$. In the case of the point $x=0$, the entropy of the rotating and non-rotating phases are equal, so no latent heat is absorbed or released at the phase transition. However, the total angular momentum still undergoes a gap, indicating a first order phase transition between the rotating and non-rotating states. The change in angular momentum is negative (the rotating phase has a negative moment of inertia), and we have
\begin{equation}\label{dJ0}
    \Delta J=J(0,\theta)=-\frac{c\sigma}{2\theta}\,.
\end{equation}

Figure \ref{fig:mu+j} shows the plots of the rotational correction of the chemical potential $\mu_r(x,\theta)$ and the specific angular momentum $J(x,\theta)$ of the rotating phase for $\theta=1$ and $A=10^{-3}$ in the range $0<x<0.3$. The plot confirms the results of the analysis and demonstrates the scales. It can be seen that the rotational correction to the chemical potential is positive for low angular velocities. It reaches its maximum for $x=x^c_1$ somewhere halfway between the phase transition points. The angular momentum curve is denoted by the blue line and is a strictly increasing function in the covered region. A part of this curve is dashed to show the metastable states. The angular momentum has zero value at the point $x=x^c_1$ and is positive for $x=x_1$. The height of the dotted line segment indicates the change in total angular momentum at the phase transition.

%... Analog of the Maxwell rule,
%\begin{equation}
%    \int J(x)dx=0.
%\end{equation}

% М: убрал пока это место в комментарии

\begin{figure}[h]
  \begin{subfigure}{.4\textwidth}
  \centering
    \includegraphics[scale=0.5]{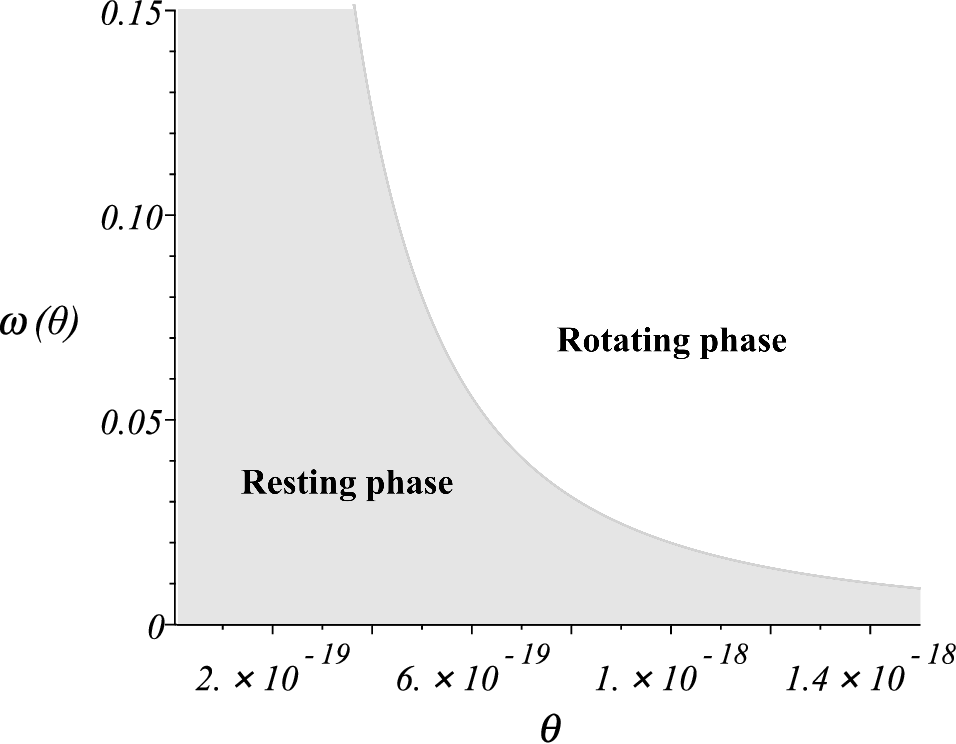}
    \caption{}
  \end{subfigure}%
  \begin{subfigure}{.6\textwidth}
  \centering
    \includegraphics[scale=0.5]{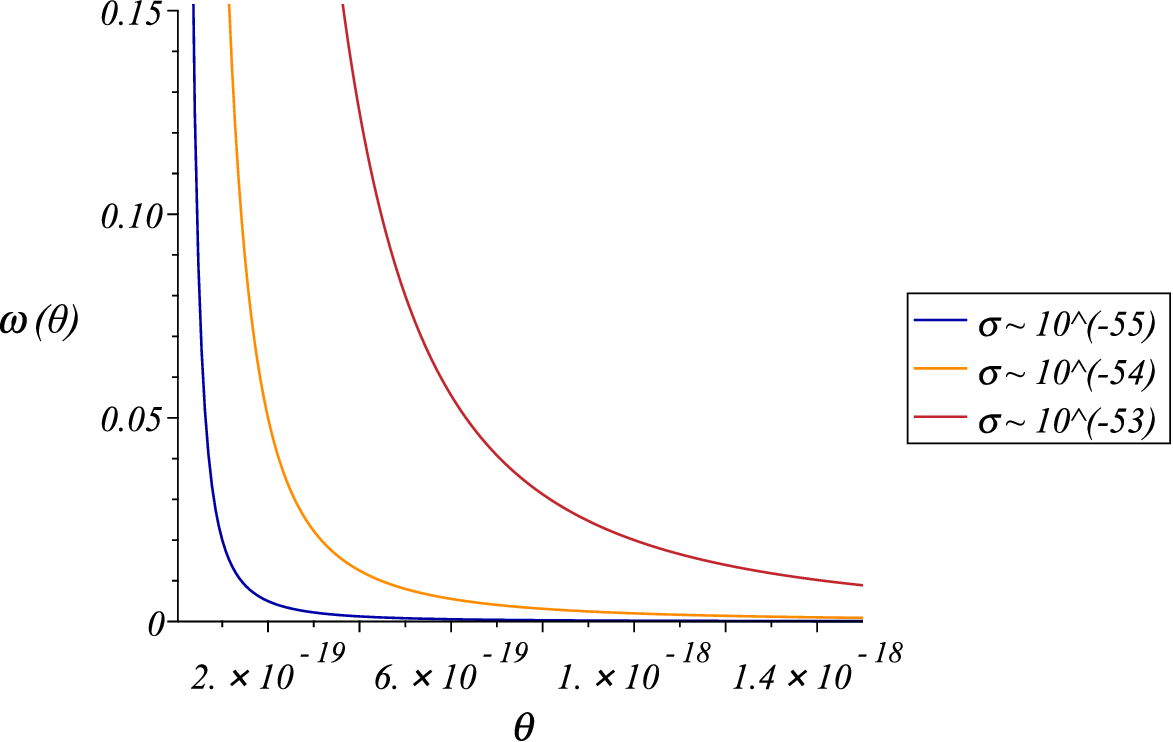}
    \caption{}
  \end{subfigure}
  \caption{Phase diagram of rotating continuous helicity gas. Figure (a) shows the diagram for $\sigma\sim10^{-53}$. Figure (b) shows phase equilibrium lines for different orders of helicity}
  \label{fig:phases}
\end{figure}
Figure \ref{fig:phases} shows the phase diagram of the continuous helicity gas. The non-rotating phase is the most stable at low angular velocities. The rotating phase is most stable at high angular velocities. The angular velocity of the phase transition decreases with increasing temperature, while decreasing temperature increases the angular velocity of the phase transition. A further decrease in temperature breaks the condition $\ln A<-1$ and the rotating phase becomes unstable for all angular velocities. Figure 2 (b) shows the dependence of the position of the phase equilibrium line on the value of the helicity. It can be seen that more extreme conditions are required for the phase transition to occur in the gas of particles with higher helicity.

\section{Conclusion}\label{sec:C}

In this paper we have studied the thermodynamics of the rotating ideal gas of classical particles with continuous helicity in three-dimensional Minkowski space. Using the results of \cite{Kaparulin2023}, we have obtained expressions for the chemical potential, angular momentum and entropy of the gas. All quantities are expressed in terms of two dimensionless parameters $x$ and $A$, which characterize the amount of rotation and the geometry of the system. The process of gas acceleration is described by a two-phase picture. The non-rotating phase is more stable at lower angular velocities, while the rotating phase is more stable at higher angular velocities. The phases are in equilibrium at two phase transition points: in the non-rotating state and at a certain value of the angular velocity, which is determined by the gas parameters. The phase transition is first-order. Although the rotating phase is less favourable at low angular velocity, it can still exist as a metastable state due to the positive specific rotation susceptibility. All these results show that interesting phase transitions can be observed in relatively simple systems such as ideal gases. Although elementary particles with continuous helicity have not been observed experimentally, the model describes the properties of ultrahot gases of massive particles with spin. We also believe that in future studies the rotating gas of particles with continuous helicity can serve as a test model to research rotationally induced phase transition phenomena in more complex thermodynamic systems.

\printbibliography
\end{document}